\documentclass[prl,epsfig,twocolumn,showpacs,preprintnumbers,amsmath,amssymb,superscriptaddress]{revtex4}
\usepackage{amsmath}
\usepackage[usenames]{color}
\usepackage{epsfig}
\usepackage{graphics}
\begin{document}

\title{Probing spatial spin correlations of ultracold gases by quantum noise spectroscopy}
\author{G. M. Bruun} 
\affiliation{Dipartimento di
Fisica, Universit\`a di Trento and CNR-INFM BEC Center, I-38050
Povo, Trento, Italy}
\affiliation{Niels Bohr Institute, University of Copenhagen,  
DK-2100 Copenhagen \O, Denmark}
\author{Brian M. Andersen}
\affiliation{Niels Bohr Institute, University of Copenhagen, 
 DK-2100 Copenhagen \O, Denmark}
\author{Eugene Demler}
\affiliation{Department of Physics, Harvard University, Cambridge, Massachusetts 02138, USA}
\author{Anders S. S{\o}rensen}
\affiliation{Niels Bohr Institute, University of Copenhagen, 
DK-2100 Copenhagen \O, Denmark}
\affiliation{QUANTOP -- Danish Quantum optics center, Niels Bohr Institute, DK-2100 Copenhagen \O, Denmark}

\date{\today{}}

\begin{abstract}
Spin noise spectroscopy with a single laser beam 
is demonstrated theoretically to provide a direct probe of the spatial correlations of cold fermionic gases. 
We show  how the generic many-body phenomena
of anti-bunching, pairing, antiferromagnetic, and algebraic spin liquid
correlations can be revealed by  measuring the  
spin noise as a function of laser width, temperature,  and frequency.

\end{abstract}


\maketitle
Ultracold atoms offer the possibility to prepare, manipulate and probe various paradigm phases of 
strongly correlated systems. Considerable efforts are devoted to 
develop sensitive detection schemes to  study  these phases. 
Whereas most experiments in this field are based on measuring mean values of various observables, further insight can be obtained from
 the correlations in the noise of the atomic distribution~\cite{Altman,folling,Hofferberth}. 
 In recent experiments a new technique using phase contrast imaging was used to probe the spin of ultracold atoms \cite{sadler06,shin06}.  
In related experiments \cite{shp,sherson,windpassinger} similar techniques have been pushed to the point where they are sensitive to the  
quantum fluctuations of the atoms.
 In this Letter, we show that quantum spin noise spectroscopy along the lines of Refs. 
 \cite{sadler06,shin06,shp,sherson,windpassinger} constitutes a 
sensitive probe of the  correlations of the underlying quantum state. 
We focus on  generic many-body phenomena such as 
antibunching, pairing,  and spin liquids. Furthermore, we show that spin noise measurement is an ideal tool for probing antiferromagnetic ordering and phase transitions for atoms in optical lattices, which is currently one of the main challenges in cold atoms physics. 
Related theoretical studies of spin noise  have recently been presented in Refs.~\cite{Eckert,mihaila,Cherng}.

Quantum noise limited probing of the spin state may be obtained either by polarization rotation \cite{shp,sherson} or phase contrast imaging  \cite{oblak,windpassinger}. In the first approach the spin imprints a phase shift on a laser beam and this phase shift is subsequently measured by interfering the beam with another laser beam (i.e. homodyne detection). In the polarization rotation measurement the two laser beams are replaced by two different polarization modes, which has the advantage that the setup is less sensitive to fluctuations in optical path length and  beam profile.
Ideally one would probe the system by imaging with a camera, but here we explore a slightly simpler situation where a laser beam is passed through the sample and the final result is measured by photo detectors without any spatial resolution. In the limit of strong beams (many photons) experiencing a small phase change, the  observable, i.e.\ the measured light quadrature in the homodyne detection,  
may be expressed as \cite{sherson,oblak,sorensen,Carusotto}
\begin{equation}
\hat{X}_{{\rm out}}=\hat{X}_{{\rm in}}+ \frac{\kappa}{\sqrt{2}} \hat{M}_z.
\label{xout}
\end{equation}
Here, $\hat{X}_{{\rm in/out}}$ is a canonical position operator describing the light normalized such that the input corresponds to vacuum noise $\langle \hat{X}_{{\rm in}}^2\rangle =1/2$, and $\kappa$ is a coupling constant.
The effective measured atomic operator is 
$
\hat{M}_z=\int d^3r \ \phi({\mathbf{r}})  \hat{s}_z({\mathbf{r}})/\sqrt{A},\label{TotalSpin}
$
where $A=\int d^3r\phi^2({\mathbf{r}})n({\mathbf{r}})/4$ is a normalization constant, $\phi({\mathbf{r}})$ is the spatial intensity profile of the laser beam,
and 
$\hat{s}_z({\mathbf{r}})=(\hat{\psi}^\dagger_\uparrow({\mathbf{r}})\hat{\psi}_\uparrow({\mathbf{r}})-\hat{\psi}^\dagger_\downarrow({\mathbf{r}})
\hat{\psi}_\downarrow({\mathbf{r}}))/2=(\hat{n}_\uparrow({\mathbf{r}})-\hat{n}_\downarrow({\mathbf{r}}))/2$ gives the local population imbalance 
(magnetization) with 
$\hat{\psi}_\sigma({\mathbf{r}})$ being the atomic field operator. We consider a two-component atomic gas ($\sigma=\uparrow,\downarrow$) with total local density 
$n({\mathbf{r}})=\langle\hat{n}_\uparrow({\mathbf{r}})+\hat{n}_\downarrow({\mathbf{r}})\rangle$ and assume Gaussian laser profiles     
$
 \phi({\mathbf{r}})\propto e^{-(x^2+y^2)/d^2}.
$
By measuring the observable $\hat X_{{\rm out}}$ it is possible to obtain spatially resolved information about the magnetization $\langle \hat X_{{\rm out}}\rangle =\kappa\langle \hat M_z\rangle/\sqrt{2}$. In many cases, however, interesting states may not have any net magnetization $\langle \hat M_z\rangle=0$. In this Letter, we will only consider  such situations and show that  a measurement of the quantum noise $\langle \hat  X_{{\rm out}}^2\rangle =(1+R \kappa^2)/2$, where  $R\equiv \langle \hat M_z^2 \rangle$ giving 
\begin{equation}
R=
\frac{1}{A}\int d^3r_1d^3r_2 \phi({\mathbf{r}}_1) \phi({\mathbf{r}}_2)\langle\hat{s}_z({\mathbf{r}}_1)\hat{s}_z({\mathbf{r}}_2)\rangle 
\label{Rdef}
\end{equation} 
provides insight into the state of the system. Since $R$ is quadratic in the atomic density operators it gives a direct measure of the atomic correlations in the system. 
The normalization in Eq. (\ref{Rdef}) is chosen such that  the quantum noise of an uncorrelated state, where each atom has an equal probability of being in each of the two internal  states, is $R=1$ (standard quantum limit).

We first consider the normal phase, where the spin fluctuations have a length scale
of $k_F^{-1}$. It follows  that $R$ vanishes if the effective volume  $V_B=(\int_V d^3r\phi)^2/\int_V d^3r\phi^2$ is large, $V_B\gg k_F^{-3}$.  Fermi statistics %(anti-bunching)
 thus suppresses the noise below the standard quantum limit  $R=1$. For a finite laser beam there will, however, be a noise contribution from the boundary $\langle M_z^2\rangle\sim d$, which translate into $R\sim 1/k_F d$.

 A key property of pairing for fermions is that the 
two particle density matrix $\langle \psi_\uparrow^\dagger(\mathbf{ r}_1)\psi_\downarrow^\dagger(\mathbf{ r}_2) \psi_\downarrow(\mathbf{ r'}_2)\psi_\uparrow(\mathbf{ r'}_1)\rangle$ has a macroscopic eigenvalue $p_cN$ with $N$ the number of particles and
 $p_c$ the condensate fraction. Spin noise spectroscopy probes the 
two particle density matrix directly, and in the large $d$ limit the noise is dominated by the largest eigenvalue  $p_cN$.
The noise depends on the shape of the applied laser beam as seen from the following argument: assuming a top hat laser profile with a radius $d$ and sharp
edges compared to the radius $\xi$ (coherence length) of the pair wavefunction $\chi(\mathbf{ r})$, the 
noise is proportional to the number of pairs within $\xi$ of the edge such that only
one particle is inside the beam. This gives a scaling  $R\propto1/ d$ as in the normal case. 
With a smooth laser profile with radius $d$ and fall-off distance $D>\xi$, the noise  is due to pairs in the edge region. These pairs couple to the gradient ($\sim1/D$) and the noise from the difference in signal from $\uparrow$  and $\downarrow$
particles is $\sim\int d^3r\chi^2(r)r^2/D^2\sim \xi^2/D^2$. This 
should be multiplied by the number of pairs in the edge region $\sim L_zdD p_c N/V$, where $L_z$ and $V$ denote
the length and volume of the system. Since $A\sim L_zd^2N/V$, 
we get $R\sim p_c  \xi^2/Dd $. With a Gaussian beam $D\sim d$, and the scaling $R\sim p_c \xi^2/d^2$ thus provides a measurement of 
$\xi$ and $p_c$.

We now use the BCS wavefunction to derive this scaling rigorously  in the BCS and BEC limits. 
Consider a homogeneous gas with  constant density $n_\sigma({\mathbf{r}})=N_\sigma/V$.  
Wicks theorem  yields
 $
\langle\hat{s}_z({\mathbf{r}}_1)\hat{s}_z({\mathbf{r}}_2)\rangle=
n\delta({\mathbf{r}})-2\theta^2(r)-2F^2(r)
$
with ${\mathbf{r}}={\mathbf{r}}_1-{\mathbf{r}}_2$,  
$\theta(r)=\langle\hat{\psi}^\dagger_\sigma({\mathbf{r}}_1)\hat{\psi}_\sigma({\mathbf{r}}_2)\rangle$,
and $F(r)=\langle\hat{\psi}_\uparrow({\mathbf{r}}_1)\hat{\psi}_\downarrow({\mathbf{r}}_2)\rangle$.
We then find
\begin{equation}
R=1-\frac{2}{A}\int d^3r_1d^3r_2 \phi({\mathbf{r}}_1) \phi({\mathbf{r}}_2) [\theta^2(r)+F^2(r)].
\label{R}
\end{equation}
In the BEC regime  $k_Fa\rightarrow0_+$, the chemical potential is $\mu\rightarrow -\hbar^2/2ma^2$. This gives 
$u_kv_k\rightarrow \Delta/2(|\mu|+k^2/2m)$ and $v_k^2\rightarrow \Delta^2/4(|\mu|+k^2/2m)^2$ for the  coherence factors defined as 
$u^2=(1+\xi/E)/2$, $v^2=1-u^2$ with $E=(\xi^2+\Delta^2)^{1/2}$ and $\xi=k^2/2m-\mu$. 
 We obtain $\theta(r)/n_\sigma=\exp(-r/a)$ and 
$F(r)/n_\sigma=\sqrt{3\pi/k_Fa}\exp(-r/a)/k_Fr$
 which is proportional to the asymptotic bound state wavefunction 
for a potential with scattering length $a$. Likewise, in the BCS limit $k_Fa\rightarrow0_-$, 
  $\theta(r)/n_\sigma=3[\sin k_Fr-k_Fr\cos k_Fr\sqrt{\pi r/2\xi}\exp(-r/\xi)]/(k_Fr)^3$ 
  and
$F(r)/n_\sigma=3\sin k_Fr\sqrt{\pi/2\xi r}\exp (-r/\xi)/k_F^2r$
 for  $r\rightarrow\infty$ where $\xi=k_F/m\Delta$ ($\hbar=1$)  and $\Delta$ is the gap. 
 Using these limiting forms in  (\ref{R}), we obtain for $d\rightarrow \infty$
 \begin{equation}
 \begin{array}{c || c | c | c}
 & \rm{Normal\: phase} & \rm{BCS\: limit} & \rm{BEC\: limit} \\\hline
 R(d)& \frac{3\pi^{1/2}}{2^{5/2}}\frac{1}{k_Fd}& \frac{\xi}{4k_Fd^{2}}& \frac{a^2}{6d^2}
\end{array}.
\label{larged}
 \end{equation}
For $s$-wave interactions, the pair wavefunction has a short-range divergence (bunching)  given by 
$F(r)=m\Delta/4\pi r$~\cite{Bruun} resulting in 
a linear decrease of the noise for $k_Fd\rightarrow 0$  in both the BCS and BEC limits. Using $p_c\sim 1/k_F\xi$, the BCS result agrees 
with the estimate given in the previous section.

\begin{figure}[t]
\includegraphics[clip=true,width=.95\columnwidth,height=0.45\columnwidth]{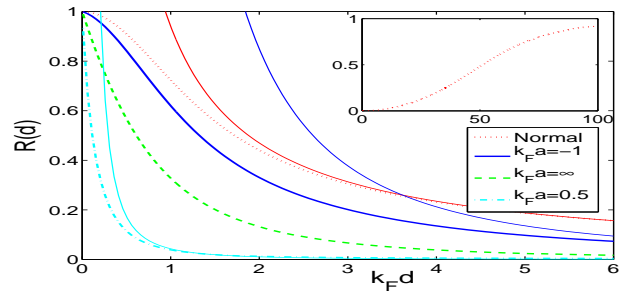}
\caption{(Color online) Noise $R(d)$ for various coupling strengths.
The thin lines show the $d\rightarrow\infty$ limit (\ref{larged}). 
  The inset shows 
the large $d$ behavior $d/L\sim{\mathcal{O}}(1)$ with $\beta=1$ for the normal phase and the superfluid phase in 
the BEC limit.}
\label{Gaussswave}
\end{figure}

In Fig. \ref{Gaussswave}, we plot $R(d)$ for a  homogeneous system of transverse 
 radius $L=100k_F^{-1}$ at $T=0$. Results for the normal phase
  and the superfluid phase with $(k_Fa)^{-1}=-1$ (BCS regime),  $(k_Fa)^{-1}=0$ (unitary limit), 
and  $(k_Fa)^{-1}=2$ (BEC regime) are shown. 
The noise is calculated numerically from (\ref{R}) using the BCS 
wavefunction. 
 The noise is \emph{below} the quantum limit $R\leq1$  and $R\rightarrow 0$ for  $d\rightarrow \infty$
in agreement with the analysis above.
Pairing suppresses the noise compared to the normal state due to 
 positive correlations between opposite spin states. 
The suppression increases with the pairing moving toward the BEC side. 

For very large $d$  the laser probes a significant fraction of the system and it is important to include possible spin fluctuations 
due to the experimental preparation of the system. Typically, such fluctuations will at least be limited by the standard quantum limit, i.e. $R=\beta$ with $\beta\gtrsim 1$  when probing the entire system. When probing  a sub-system this gives an extra contribution 
 $\sim \beta V_B/V$ which is important for large $d$  for the normal phase and the superfluid phase 
 on the BEC side $(k_Fa)^{-1}=2$ (see inset in Fig. \ref{Gaussswave}). 
 However, this term is absent for $(k_Fa)^{-1}\lesssim0.5$,  
   since superfluidity quenches the spin noise in this regime~\cite{Shin,Pilati,NoiseEdge}. Observing $R\ll 1$ for a large portion of the sample would represent an extreme  experimental demonstration of this quenching. 

The observed enhancement of the nuclear spin relaxation just below the transition temperature $T_c$
(Hebel-Slichter effect) constitutes one of the hallmark experimental tests of BCS theory.
 We now demonstrate the existence of a spin noise spectroscopy analogy 
to the Hebel-Slichter effect. Similar effects has been demonstrated to occur in inelastic light scattering 
and Bragg scattering experiments~\cite{BruunSlichter}. 
The probing technique discussed in this Letter
 is in principle non-destructive. By recoding the signal for a long duration of time one can thus obtain all frequency components of the noise $R(d,\omega)$ by Fourier analysis~\cite{mihaila}, i.e. Fourier transforming  the measured $\hat X_{{\rm out }}(t)$ provides a measurement of $M_z(\omega)$. Such probing will have similar signal-to-noise ratio $\kappa^2(
\omega)\sim \eta\alpha$, but since  spontaneous emission may lead to significant heating  $\eta$ may have to be kept very low to avoid that the system heats up during the measurement. 
Using  (\ref{R}) we obtain 
for a homogeneous system
\begin{gather}
R(d,\omega)=\frac{8\pi md^2}{n}\int \frac{d^3k}{(2\pi)^3}
\frac{E'}{\sqrt{E'^2-\Delta^2}}(uu'+vv')^2\nonumber\\
\times f(1-f')e^{-({\mathbf{k}}_\perp-{\mathbf{k}}_{\perp}')^2d^2/2}I_0(k_\perp k'_\perp d^2)
\label{Slichter2}
\end{gather}
where $I_0$ is the modified Bessel function of the first kind,
 ${\mathbf{k}}_\perp=(k_x,k_y)$ is the transverse momentum, and  $f=[\exp(\beta E)+1]^{-1}$. 
The primed quantities refer to the momentum $\mathbf{k}'$ with energy  $E'=E+\omega$. There 
is momentum conservation along the $z$-direction with  $k_z'=k_z$ 
whereas ${\mathbf{k}}'_\perp\neq {\mathbf{k}}_\perp$  due to the transverse Gaussian profile.
 Eq. (\ref{Slichter2}) gives the noise contribution from  
quasiparticle scattering from momentum ${\mathbf{k}}$ to ${\mathbf{k}}'$. 
There are additional terms describing 
pair breaking and quasiparticle absorption which do not affect the Hebel-Slichter effect.  

\begin{figure}
\includegraphics[width=0.8\columnwidth,height=0.45\columnwidth]{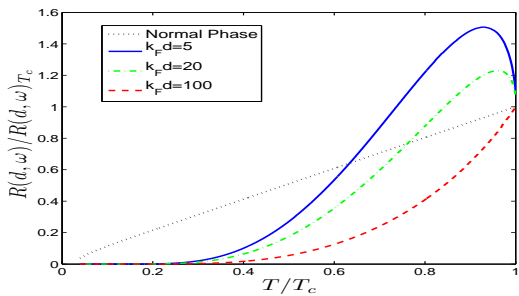}
\caption{(Color online) $R(d,\omega)$ in units of $R(d,\omega)$ at $T_c$ for various laser widths $d$ and $k_F a=-1$.}
\label{SlichterFig}
\end{figure}

In Fig.\ \ref{SlichterFig}, we plot $R(d,\omega)$ as a 
function of $T/T_c$ calculated numerically from (\ref{Slichter2}) using the self-consistently determined gap
$\Delta(T)$. We have chosen $k_Fa=-1$ giving $T_c/T_F\simeq0.13$ and 
$\omega/k_BT_c=0.08$ since the Hebel-Slichter effect only 
occurs for $\omega\lesssim k_BT_c$. For narrow laser widths, a
Hebel-Slichter peak is prominent below $T_c$. The peak arises from an
increased density of states at the gap edge; it 
decreases with increasing $d$  and disappears for $d\gg \xi(T=0)\simeq 9$. This is because for large laser widths, 
the scattering becomes
subject to momentum conservation which restricts the available phase space.

Quantum systems in periodic potentials constitute 
another class of intriguing systems which can be examined by cold atomic gases using optical lattices.
Superfluidity in lattices, possibly of $d$-wave symmetry,
 can be detected by suppression of spin noise similar to the discussion above 
for homogeneous systems. The only difference is that the $d$-wave pair wavefunction 
does not diverge for short length scales and there is no linear decrease in $R$ for small laser radii $d$.
One could use a laser with elliptical transverse profile to detect the anisotropic suppression of spin noise due  to the $d$-wave symmetry of the pairing. 

\begin{figure}[t]
\begin{center}
\leavevmode
\begin{minipage}{.49\columnwidth}
\includegraphics[clip=true,height=1.06\columnwidth,width=1\columnwidth]{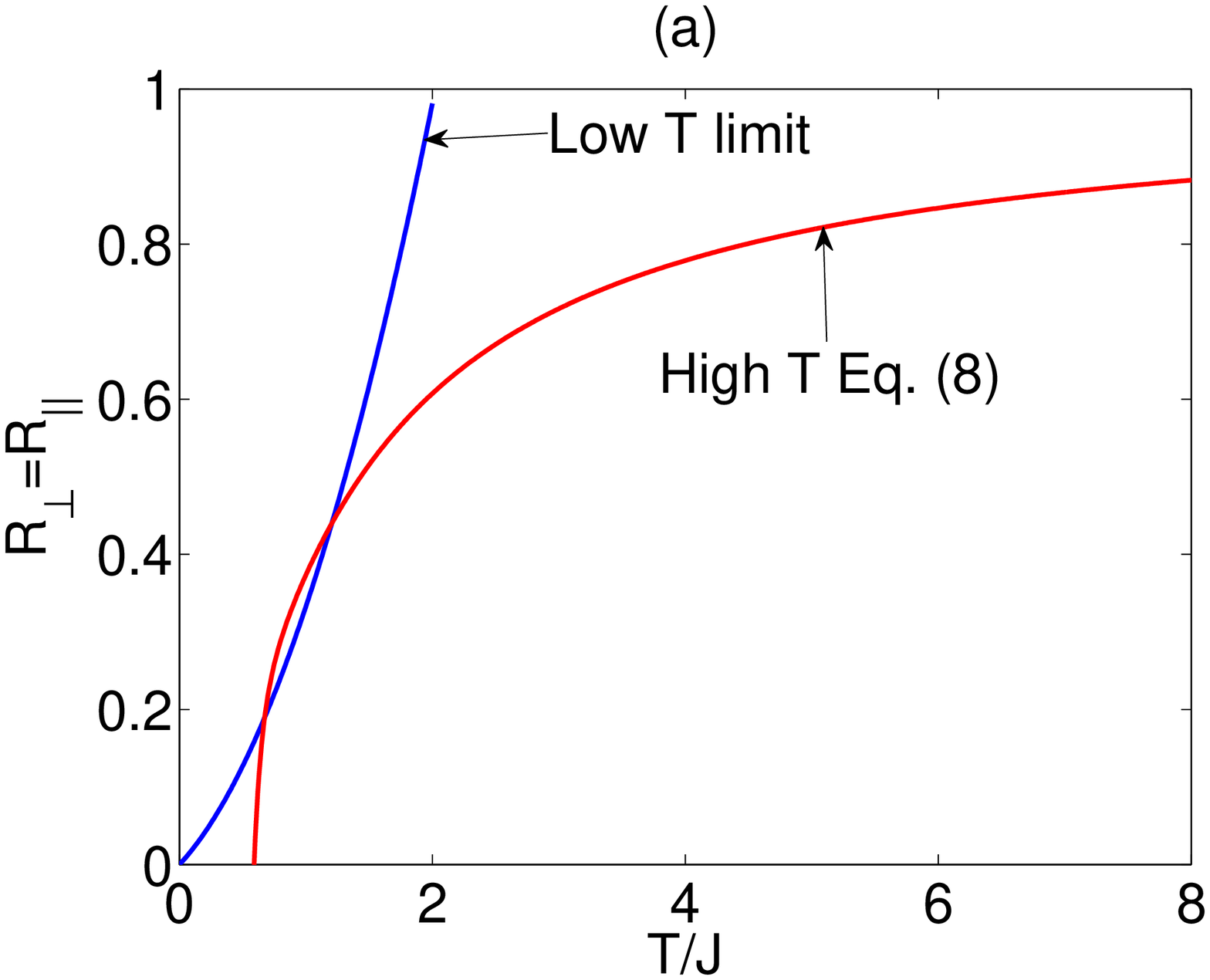}
\end{minipage}
\begin{minipage}{.49\columnwidth}
\includegraphics[clip=true,height=1.06\columnwidth,width=1\columnwidth]{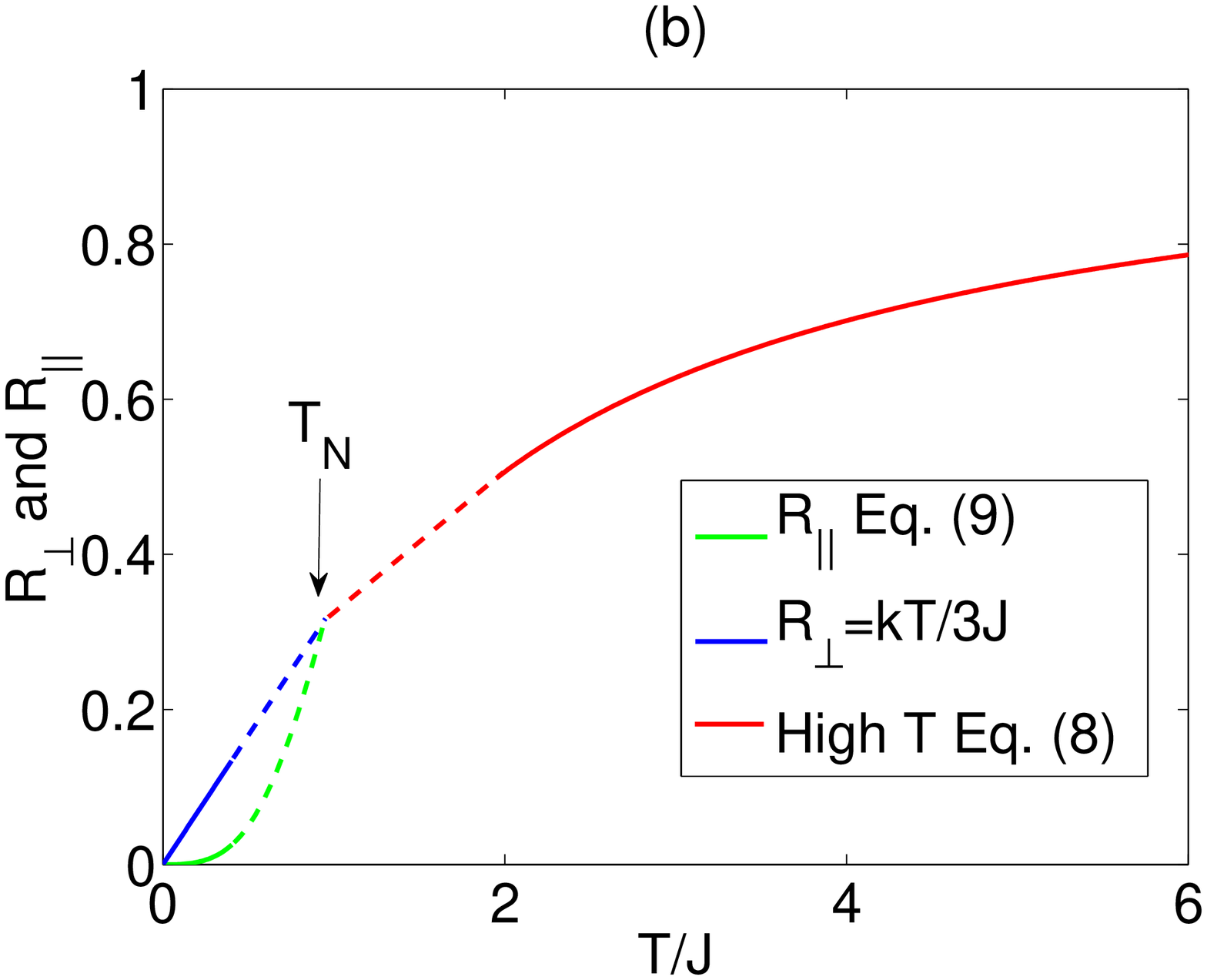}
\end{minipage}
\caption{(Color online) $R_\parallel$ and $R_\perp$ for a 2D (a) and 3D (b) system. Solid lines are the high and low $T$ results discussed in the text 
and the dashed lines in (b) (obtained by 
simple rescaling) indicate how they meet at $T_N\simeq0.946J$~\cite{Sandvik}.}
\label{LatticeFig}
\end{center}
\end{figure}

Presently, a main experimental goal in optical lattices is to observe the onset of antiferromagnetic (AF) correlations 
with decreasing temperature~\cite{Jordens}. As demonstrated below, spin noise spectroscopy can measure the magnetic susceptibility of the system and hence constitutes an important experimental probe of the spin correlations. As an example, we study atoms described by the Hubbard model which  
in the strong repulsion limit at half filling for $kT\ll U$ reduces to the AF  Heisenberg model, 
$H=J\sum_{\langle i,j \rangle}\textbf{s}_i\cdot\textbf{s}_j$,
where  $\langle i,j \rangle$ denotes nearest neighbor pairs and $\textbf{s}_i$ is the spin $1/2$ operator
for the atoms at site $i$. Assuming, without loss of generality, a staggered magnetization along the $z$-direction, we now 
show how to detect AF correlations by measuring $R_\parallel\equiv\langle \hat M_z^2 \rangle$ and $R_\perp\equiv\langle \hat M_x^2 \rangle$,  where $M_x$ is defined  analogous to $M_z$. (A preferred direction for the broken symmetry can be induced by enforcing a slight anisotropy in the exchange coupling $J$.) Here we are mainly interested in the $T$ dependence and focus on the situation where we probe the entire ensemble. 
Therefore, we assume a broad laser profile with $\phi=1$ in (\ref{Rdef}) such that  $R_{\parallel(\perp)}=4\langle S_{z(x)}S_{z(x)}\rangle/N$ with  ${\mathbf{S}}=\sum_i {\mathbf{s}}_i$ and $N$ is the number of spins. In the paramagnetic phase, $R_\parallel=R_\perp=4kT\chi$ where $\chi$ is the magnetic susceptibility.
A high temperature expansion yields for the 
2D square and 3D cubic lattices~\cite{DombGreen}
\begin{equation}
4kT\chi=\left\{\begin{array}{lcl}
1-2x+2x^2-1.333x^3+\ldots  &,&\rm{2D}\\
1-3x+6x^2-11x^3+\ldots  &,&\rm{3D}
\end{array}
\right.
\label{HighT}
\end{equation}
where  $x=J/2kT$. 
In 2D, the system remains paramagnetic for $T>0$, and modified spin-wave theory yields $\chi=(12J)^{-1}[0.524+0.475T/J+{\mathcal{O}}(T^3)]$
for $T/J\ll 1$~\cite{Takahashi}. In the 3D case, the system undergoes a phase transition to an AF phase at the N\'{e}el 
temperature $T_N$. 
In the AF phase, $R_\parallel\neq R_\perp$. Using spin-wave theory for $T<T_N$, we obtain 
$R_\perp=kT/(3J)$  and
\begin{equation}
R_\parallel=\frac{4}{N}\sum_{\mathbf{k}}\frac{1}{2\sinh^2(\beta\omega_{\mathbf{k}}/2)}.
\label{SzSz3D}
\end{equation} 
Here $\omega_{\mathbf{k}}=3J\sqrt{1-\gamma_{\mathbf{k}}^2}$ is the spin-wave energy with $\gamma_{\mathbf{k}}=(\cos k_xa+\cos k_ya+\cos k_za)/3$
for a cubic lattice with lattice constant $a$. 
 The sum in (\ref{SzSz3D}) is over the reduced Brillouin zone. For $kT\ll J$, (\ref{SzSz3D}) yields
$R_\parallel=4(kT)^3/(3s^3)$, with $s=\sqrt{3}J$ the spin-wave velocity. 
 In Fig.\  \ref{LatticeFig} we plot these results for  both the 2D and 3D systems. 
We see that the onset of AF correlations in the paramagnetic phase can be detected as a decrease in the noise from the uncorrelated result 
$R_\parallel=R_\perp=1$ as described by (\ref{HighT}). By comparing with the  high temperature expansion, the spin noise may even serve as an accurate thermometer for the spin temperature.
Furthermore, the AF phase for the 3D case can be detected by observing $R_\parallel\neq R_\perp$.  
An advantage of probing collective operators like $S_z$ is that they are conserved, and therefore could be measured after time of flight. 
In this case, however, special care has to be taken of the contribution from the boundary region.
We do not expect the trapping potential to change these results qualitatively~\cite{Andersen,snoek}.

The method presented here can also be used to probe the correlations of more exotic quantum phases such as 
resonating valence bond states and algebraic spin liquids \cite{hermele}. These states are characterized  by long-range spin correlations $\langle {\mathbf{s}}({\mathbf{r}}){\mathbf{s}}(0)\rangle\sim  (-1)^{r_x+r_y}/ r^{(1+\eta)}$.
Techniques exist for addressing, e.g., every second site in a lattice \cite{peil}. Flipping every second spin before the measurement ($s_z\rightarrow (-1)^{r_x+r_y}$) will give 
$\langle s_z({\mathbf{r}})s_z(0)\rangle\sim  1/r^{(1+\eta)}$. Performing noise spectroscopy on this state will give a contribution from the long-range correlations  $R\sim d^{(1-\eta)}$. By measuring the scaling of $R$ with $d$ one can thus directly determine the exponent $\eta$ of the spin correlations.

Finally we consider the experimental requirements for realizing our scheme. The experiments should be quantum noise limited with all classical noise 
sources suppressed. This has already been achieved in several experiments \cite{shp,sherson,windpassinger,oblak}, and we expect it to be simpler to realize for the 
smaller systems considered here.
 In addition,  the atomic noise should be large compared to the light noise inherently present in the probe. 
 The spontaneous emission probability pr.\ atom  caused by the probing light is 
 $\eta\sim\kappa^2 /\alpha$, where $\alpha=3nL_z\lambda^2\gamma_x/\gamma 2\pi$
  is the optical depth of the ensemble \cite{sherson,oblak,sorensen}. 
 Taking $n\sim10^{12}$ cm$^{-3}$, $L_z\sim 100$ $\mu$m,  a probing wavelength $\lambda=671$ nm corresponding to Li, and a branching ratio $\gamma_x/\gamma=1/2$ gives $\alpha=16$ for a harmonically trapped Fermi gas. For atoms in optical lattices at half filling  $\alpha\approx N_s=50$ where $N_s$ is the number of lattice sites in each direction. One can thus have a large signal-to-noise ratio $\kappa^2R\gtrsim 1$ with very little noise added from spontaneous emission during the probing 
 $\eta=\kappa^2/\alpha\ll1$. Another concern
 is the spatial resolution. Experimentally, one may obtain a resolution down to $d\sim 5 \lambda$ \cite{shin06}.  Taking $n\sim 10^{12}$cm$^{-3}$  this 
corresponds to $k_F d\sim 10$.   Thus, it may require an adiabatic expansion of the gas to
observe the small scale limit of Fig. \ref{Gaussswave}. However, it is possible directly to observe the large $d$ scaling, the Hebel-Slichter effect, and the onset of magnetic correlations. 
 
In summary, we have shown how to extract the correlations of quantum states of ultracold atoms using
spin noise spectroscopy. This was demonstrated explicitly by calculating the spin noise for normal Fermi gasses, superfluids, paramagnetic and AF phases and algebraic spin liquids. 
This method can be applied to other strongly correlated systems as well as extended  to higher order moments \cite{Cherng}. 
It may even be extended to full quantum state tomography of the two particle density matrix.
 
 We acknowledge useful conversations with R.\ Cherng, E.\ Polzik, A.\ Sanpera. Partial support was provided by  the Villum Kann Rasmussen Foundation (B.\ M.\ A.) and the Harvard-MIT CUA, DARPA, MURI, and the NSF grant
DMR-0705472 (E.\ D.).


\begin{thebibliography}{99}
\bibitem{NoiseEdge} We assume the laser does not probe the far edges of the gas
where any spin imbalance exists as a normal phase.
\end{thebibliography}
\end{document}